\documentclass[twocolumn,aps,prl,epsfig,superscriptaddress,showpacs]{revtex4}

\usepackage{graphicx,amsmath,amsfonts,textcomp}

\begin{document}

\title{Multiferroic phase transition near room temperature in BiFeO$_{3}$ films}
\author{I.C. Infante}
\affiliation{Unit\'e Mixte de Physique CNRS/Thales, Campus de l'Ecole Polytechnique, 1 avenue Fresnel, 91767 Palaiseau, France and Universit\'e Paris-Sud, 91405, Orsay, France }
\affiliation{Laboratoire Structures, Propri\'et\'es et Mod\'elisation des Solides, UMR 8580 CNRS-Ecole Centrale Paris, Grande Voie des Vignes, 92295 Ch\^atenay-Malabry Cedex, France}
\author{J. Juraszek}
\affiliation{Groupe de Physique des Mat\'eriaux, UMR6634 CNRS-Universit\'e de Rouen, F-76801 St. Etienne du Rouvray, France}
\author{S. Fusil}
\affiliation{Unit\'e Mixte de Physique CNRS/Thales, Campus de l'Ecole Polytechnique, 1 avenue Fresnel, 91767 Palaiseau, France and Universit\'e Paris-Sud, 91405, Orsay, France }
\author{B. Dup\'e}
\affiliation{Laboratoire Structures, Propri\'et\'es et Mod\'elisation des Solides, UMR 8580 CNRS-Ecole Centrale Paris, Grande Voie des Vignes, 92295 Ch\^atenay-Malabry Cedex, France}
\affiliation{CEA-DAM DIF, F-91297 Arpajon, France}
\author{P. Gemeiner}
\affiliation{Laboratoire Structures, Propri\'et\'es et Mod\'elisation des Solides, UMR 8580 CNRS-Ecole Centrale Paris, Grande Voie des Vignes, 92295 Ch\^atenay-Malabry Cedex, France}
\author{O. Di\'eguez}
\affiliation{Institut de Ci\`encia de Materials de Barcelona (ICMAB-CSIC), Campus UAB, E-08193 Bellaterra, Spain}
\author{F. Pailloux}
\affiliation{Institut Pprime, UPR 3346 CNRS-Universit\'e de Poitiers-ENSMA, SP2MI, 86962 Futuroscope-Chasseneuil Cedex, France}
\affiliation{Canadian Center for Electron Microscopy, Brockhouse Institute for Materials Research, McMaster University, 1280 Main Street West, Hamilton, Ontario L8S4M1, Canada}
\author{S. Jouen}
\affiliation{Groupe de Physique des Mat\'eriaux, UMR6634 CNRS-Universit\'e de Rouen, F-76801 St. Etienne du Rouvray, France}
\author{E. Jacquet}
\affiliation{Unit\'e Mixte de Physique CNRS/Thales, Campus de l'Ecole Polytechnique, 1 avenue Fresnel, 91767 Palaiseau, France and Universit\'e Paris-Sud, 91405, Orsay, France }
\author{G. Geneste}
\affiliation{Laboratoire Structures, Propri\'et\'es et Mod\'elisation des Solides, UMR 8580 CNRS-Ecole Centrale Paris, Grande Voie des Vignes, 92295 Ch\^atenay-Malabry Cedex, France}
\affiliation{CEA-DAM DIF, F-91297 Arpajon, France}
\author{J. Pacaud}
\affiliation{Institut Pprime, UPR 3346 CNRS-Universit\'e de Poitiers-ENSMA, SP2MI, 86962 Futuroscope-Chasseneuil Cedex, France}
\author{J. \'I\~niguez}
\affiliation{Institut de Ci\`encia de Materials de Barcelona (ICMAB-CSIC), Campus UAB, E-08193 Bellaterra, Spain}
\author{L. Bellaiche}
\affiliation{Physics Department, University of Arkansas, AR 72701 Fayetteville, USA}
\author{A. Barth\'el\'emy}
\affiliation{Unit\'e Mixte de Physique CNRS/Thales, Campus de l'Ecole Polytechnique, 1 avenue Fresnel, 91767 Palaiseau, France and Universit\'e Paris-Sud, 91405, Orsay, France }
\author{B. Dkhil}
\affiliation{Laboratoire Structures, Propri\'et\'es et Mod\'elisation des Solides, UMR 8580 CNRS-Ecole Centrale Paris, Grande Voie des Vignes, 92295 Ch\^atenay-Malabry Cedex, France}
\author{M. Bibes}
\email{manuel.bibes@thalesgroup.com} \affiliation{Unit\'e Mixte de Physique CNRS/Thales, Campus de l'Ecole Polytechnique, 1 avenue Fresnel, 91767 Palaiseau, France and Universit\'e Paris-Sud, 91405, Orsay, France }

\begin{abstract}

\vspace{0.2cm}

In multiferroic BiFeO$_3$ thin films grown on highly mismatched LaAlO$_3$ substrates, we reveal the coexistence of two differently distorted polymorphs that leads to striking features in the temperature dependence of the structural and multiferroic properties. Notably, the highly distorted phase quasi-concomitantly presents an abrupt structural change, transforms from a hard to a soft ferroelectric and transitions from antiferromagnetic to paramagnetic at 360$\pm$20 K. These coupled ferroic transitions just above room temperature hold promises of giant piezoelectric, magnetoelectric and piezomagnetic responses, with potential in many applications fields.

\vspace{0.2cm}

\end{abstract}


\maketitle

Multiferroics that display simultaneously magnetic, polar and elastic order parameters, are gaining much attention due to their fascinating fundamental physics \cite{wang2009a} as well as their considerable application potential \cite{bea2008a,bibes2011}. Among them, BiFeO$_3$ (BFO) is intensively studied because both ferroelectric (FE) and magnetic orders coexist at room temperature (RT) \cite{catalan2009}. Below the Curie temperature T$_C \approx 1100$ K, BFO is described by the rhombohedral R3c space group, which allows antiphase octahedral tilting and ionic displacements from the centrosymmetric positions about and along the same $<$111$>$ cubic-like direction. Bulk BFO is also a G-type antiferromagnet (AF) with a cycloidal spin modulation below the N\'eel temperature T$_N$=640K.

The coexistence of ferroic orders with several lattice instabilities makes BFO an interesting playground to investigate strain engineering \cite{infante2010,hatt2010,dieguez2011}. In particular, a novel "T-like" phase with giant tetragonality \cite{ricinschi2006,bea2009,christen2011} and enhanced properties \cite{zeches2009} was revealed for misfit strains of $\sim$-5\%. Also, and in contrast with the situation in more "classical" FEs like BaTiO$_3$ \cite{choi2004} it was found that compressive strain depresses T$_C$ while leaving T$_N$ almost unchanged, as a consequence of the subtle interplay between polarization and oxygen octahedral tilts \cite{infante2010}. For compressive strain values lower than -2.5\% T$_C$ and T$_N$ should meet \cite{infante2010}, extending the interest of BFO films as the magnetoelectric response should then be enhanced \cite{wojdel2010,prosandeev2011}.

Here we show that AF and FE phase transitions occur quasi-concomitantly in BFO films deposited onto LaAlO$_3$ (misfit strain of -4.8\%). Remarkably, this multiferroic phase transition takes place just above RT, in a range of interest for applications. Combining temperature-dependent X-ray diffraction (XRD), piezoresponse force microscopy (PFM), M\"{o}ssbauer (MS) and Raman spectroscopy techniques, we evidence that both transitions occur in the temperature domain of 340 K - 380 K. These findings are supported by theoretical calculations based on first-principles.

\begin{figure}
\includegraphics[width=0.95\columnwidth]{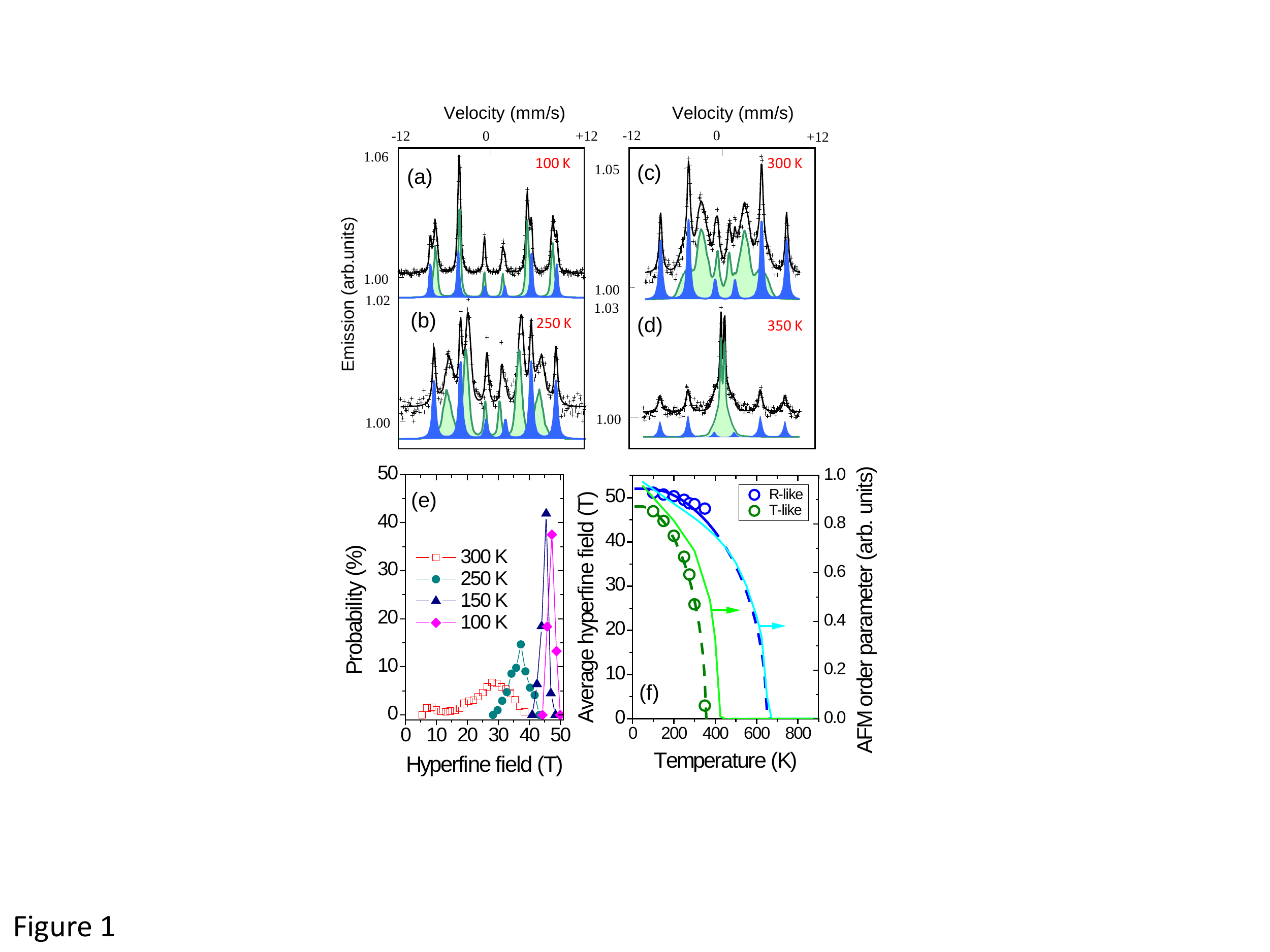}
\caption{(color online) $^{57}$Fe M\"{o}ssbauer spectra of the BFO//LAO film at 100 K (a), 150 K (b), 250K (c) and 300 K (d) fitted by two magnetic sub-spectra (blue : R-like and green : T-like). Hyperfine field distributions of the T-like component at different temperatures (e) and temperature dependence of the average hyperfine fields (f). Dashed lines are Brillouin functions for Fe$^{3+}$ (S=5/2) fitted to the data; solid lines correspond to model Hamiltonien simulations.}
\label{moss}
\end{figure}

\begin{figure}
\includegraphics[width=0.95\columnwidth]{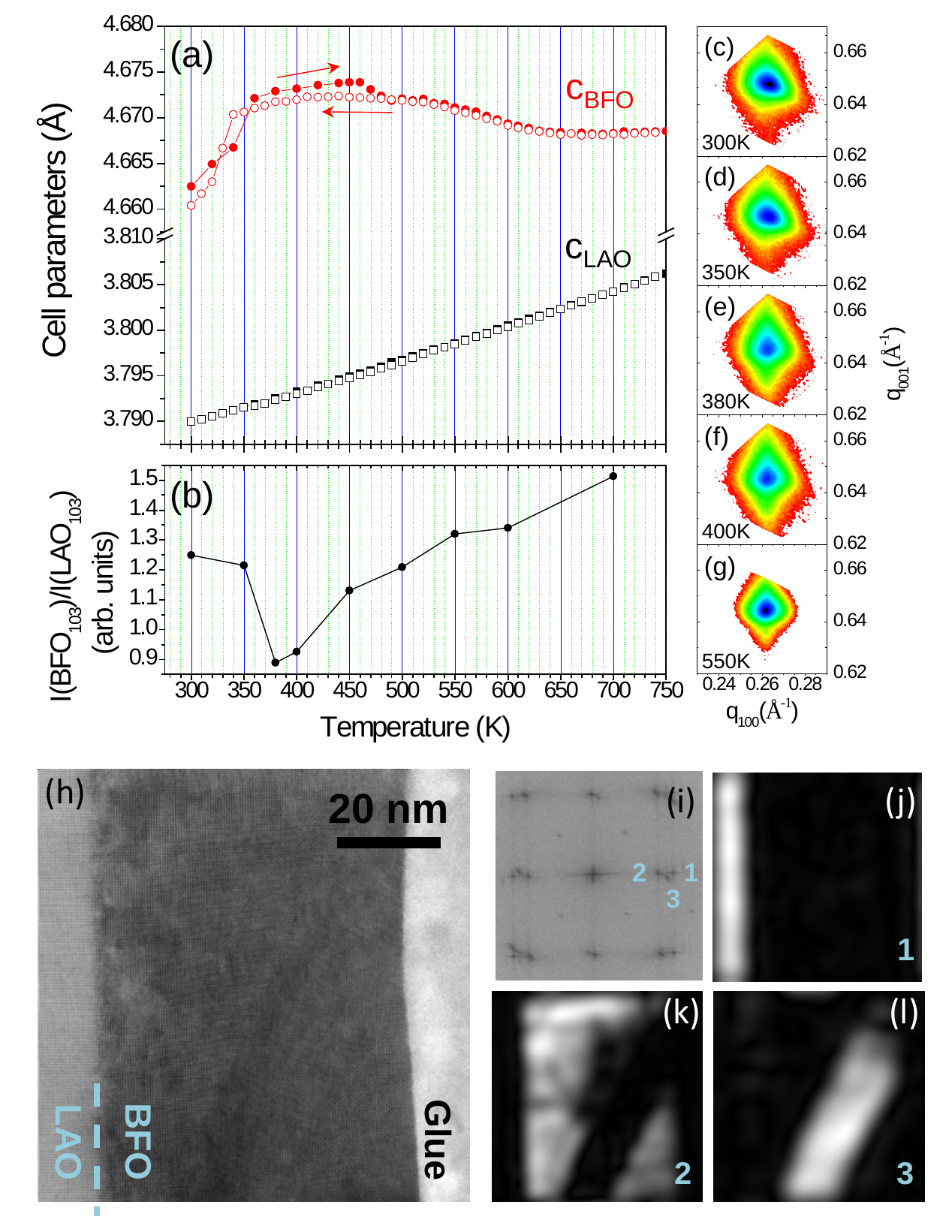}
\caption{(color online) (a) Temperature dependence of the c axis parameters of the T-like phase of the BFO film and of the LAO substrate. (b) Temperature dependence of the relative intensity of the (103) reflections from the T-like BFO phase and the LAO. (c-g) Reciprocal space maps of the (103) reflection of the T-like BFO phase for different temperatures. (h) Bright field image of a BFO film, (i) FT of (h) showing spots coming from the different (labeled 1,2 and 3), (j-l) digital dark field images of the considered spots. }
\label{struc}
\end{figure}

The films were grown by pulsed laser deposition in conditions reported elsewhere \cite{bea2005,bea2009}. The 70 nm-thick sample used for MS was grown using a $\sim$100\% $^{57}$Fe enriched target. For PFM measurements, an 11 nm-thick fully strained La$_{2/3}$Sr$_{1/3}$MnO$_3$ (LSMO) bottom electrode was used. Symmetric X-ray diffraction (XRD) 2$\theta - \omega$ scans indicated a majoritary phase with a large c axis parameter of 4.67 \AA (corresponding to the T-like phase, but crystallizing in a Cm space group, as inferred from high-resolution scanning transmission electron microscopy, HRSTEM), while asymmetric scans indicated the presence of an additional R-like phase. This phase coexistence is confirmed by HRSTEM, see Fig. \ref{struc}h-l. Fig. \ref{struc}h displays a bright field image and Fig. \ref{struc}i the corresponding diffraction pattern. Digital dark field images were calculated from the Fourier transform (FT) of the HRSTEM image by selecting a spot of a given phase and performing an inverse FT. In this FT, three families of spots can be identified along the growth direction: the LAO substrate (Fig. \ref{struc}j), T-like BFO with a large c/a ratio (Fig. \ref{struc}k) and R-like BFO with the c-axis tilted of about 3 degrees from the growth direction (Fig. \ref{struc}l). The R-like diffraction can then be attributed to the large slanted grain and the main T-like phase is visible in the rest of the image. Using the GPA method, the average out-of-plane and in-plane lattice parameters for the T-like and R-like phases are c=4.66 \AA and a=3.79 \AA, and c=4.10 \AA and a=3.91 \AA, respectively. The R-phase thus experiences a -1.3\% compressive strain, as in BFO films grown on SrTiO$_3$ \cite{infante2010}.

$^{57}$Fe M\"{o}ssbauer spectra were measured using the conversion electron technique (CEMS) which allows the characterization of thin films \cite{juraszek2009}. The spectra were collected under normal incidence using a gas-flow proportional counter mounted inside a closed cycle Janis cryostat, and a $\sim$50 mCi $^{57}$Co radioactive source in a Rh matrix in constant acceleration mode. The isomer shift is given with respect to $\alpha$-Fe at 300 K. Fig. \ref{moss}a-d displays the spectra at different temperatures. The data unambiguously show the contribution of two AF components associated with two different sets of magnetically split sextets between 100 K and 300 K. Isomer shifts values ($\delta$=0.31-0.37 mm/s) for both components are characteristic of Fe$^{3+}$ ions in octahedral coordination. At 100 K, the spectrum exhibits a sharp sextet with a magnetic hyperfine field B$_{hf}$ = 51 T and a slightly broader one fitted with a narrow magnetic hyperfine field distribution P(B$_{hf}$) centered at B$_{hf}$ =47 T. The sharp sextet has a B$_{hf}$ close to that of bulk BFO and is identified to the R-like phase whereas the less magnetic component is identified to the T-like phase. Both sub-spectra are very similar to the spectrum recorded for BFO grown on GdScO$_3$ (GSO) \cite{infante2010} indicating that low temperature antiferromagnetism is G-type. It should be noted that the line intensity ratio R$_{23}$ of the 2nd and 3rd lines of each Zeeman sextet is close to 4.0, evidencing the in-plane orientation of the Fe$^{3+}$ magnetic moments for both phases.

The spectral area ratio between both magnetic phases, represents roughly 2/3 (T-like) and 1/3 (R-like) of the sample volume at all temperatures, meaning that one phase does not transform into the other. In fact, the coexistence of both phases is not an unexpected feature considering their energy diagrams as a function of misfit strain \cite{dupe2010,chen2011}. We argue that the minority-phase allows the stabilization of the giant tetragonality majority-phase. Indeed, the T-phase is associated with a misfit-strain of -4.8\%, whereas the addition of 1/3 of R-phase with a misfit-strain of -1.3\% would decrease the global misfit-strain felt by the BFO film to $\sim$-3.6\%.

Upon increasing temperature up to 300 K, the overall splitting of the outer emission lines, that measures the average effective B$_{hf}$ at the nucleus, drastically decreases for the T-like component, revealing the approach of a magnetic phase transition (Figs. \ref{moss}b-d). The corresponding P(B$_{hf}$) distribution becomes broader and its maximum is progressively shifted to lower fields (Fig. \ref{moss}e). The temperature dependence of the average B$_{hf}$ for both the T-like and R-like phases, and their fits using a mean field model are shown in Fig. \ref{moss}f. For the T-phase they indicate a T$_N\approx$360 K far from the bulk value of 640 K. The same analysis for the R-like phase yields a T$_N$ of $\approx$640 K as expected from the calculated phase diagram based on a monoclinic Cc ground state \cite{infante2010}.

The proximity of the T$_N$ for the T-phase with RT is also confirmed by our simulations. It has been shown \cite{dieguez2011b} that the magnetic interactions in BFO super-tetragonal phases are characterized by a strong splitting of the in-plane (J$_{ab}$) and out-of-plane (J$_c$) couplings between nearest-neighbouring Fe spins. Further, the basic magnetic properties can be captured in a simple Heisenberg Hamiltonian (E=E$_0$+\textonehalf J$_{ij}$S$_i$S$_j$, with $|$S$_i|$=1 and the sum being limited to first-nearest neighbors) with J$_{ab}\approx$40 meV and J$_c\approx$4 meV. We solved such a model using standard Montecarlo techniques and obtained the temperature dependence of the AF order parameter shown with solid lines in Fig. \ref{moss}f; the computed T$_N$ is about 425 K, in reasonable agreement with our experimental observation for the T-like phases. (We checked that the computed T$_N$ remains essentially unchanged upon $\pm$2 meV variations of the coupling constants.) In Fig. \ref{moss}f we also show the simulation results for the analogous Heisenberg Hamiltonian corresponding to BFO's bulk-like phase (defined by J$_{ab}$ = J$_c$ = 38 meV); in this case we obtained T$_N \simeq$650 K, in good agreement with our experimental value of 640 K for the R-like phase. Our theory thus confirms that the majority T-like and minority R-like phases in our films present a markedly different temperature dependence of their magnetic properties, as well as distinct magnetic critical temperatures. From this analysis, we conclude that the films consist of a mixture of a 1.3\% strained R-like phase, whose physical properties are described in Ref. \cite{infante2010}, and a poorly known T-like phase that we discuss further in the following.

\begin{figure}
\includegraphics[width=0.9\columnwidth]{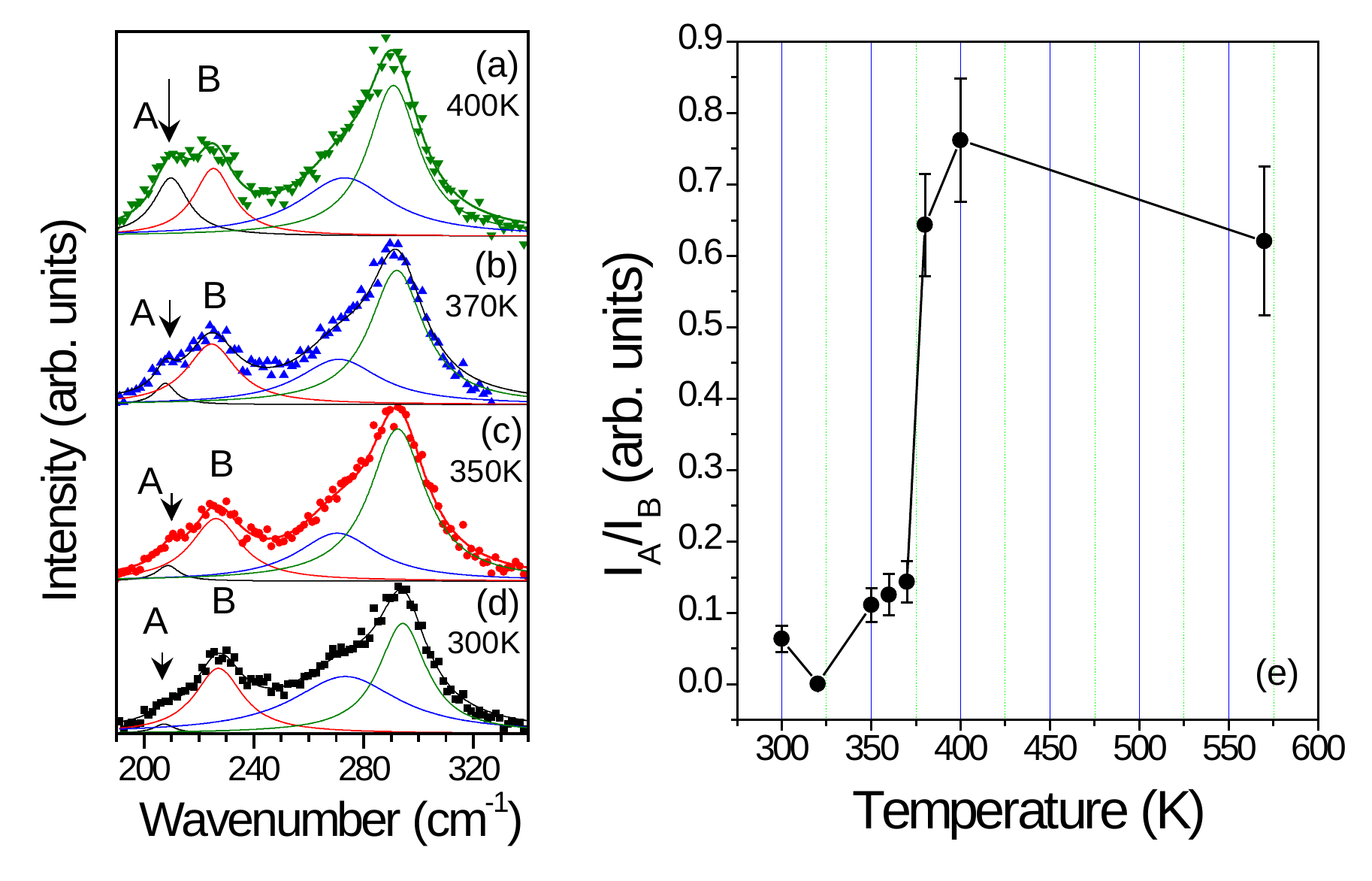}
\caption{(color online) (a-d) Raman spectroscopy data at different temperatures (symbols) along with fits (lines) of the different peak components. (e) Relative intensity of the peaks labeled "A" and "B" in (a-d) as a function of temperature.}
\label{raman}
\end{figure}

We first characterized the FE properties as a function of temperature using XRD \cite{infante2010,toupet2010}. Fig. $\ref{struc}$a displays the thermal variation of the c-axis parameters of the T-like phase and the substrate. The c-axis parameter increases rapidly as temperature rises before reaching a plateau above $\sim$340 K (heating) or $\sim$370 K (cooling). The plateau spreads over 150 K before the c-axis parameter starts to decrease continuously upon increasing temperature. While another change of behaviour is visible at $\sim$700 K and might be evidence for a structural modification, the abrupt change at 340 K-370 K appears as a signature of a reversible structural phase transition that is clearly of first order. As the elastic change is rather strong, it cannot be directly due to the AF to paramagnetic phase transition that we evidenced using MS \cite{infante2010,toupet2010,haumont2008}. More likely, this transition is related to a FE phase transition. More insight on this point is gained from the temperature dependence of the (103) reflection. As visible in Fig. $\ref{struc}$b, its normalized intensity shows an obvious change of behavior at $\sim$380K. Its shape also changes from asymmetric (Fig. $\ref{struc}$c), in agreement with a monoclinic-like phase \cite{noheda2002,christen2011}, to symmetric above 370 K (Fig. $\ref{struc}$e-g), which strongly suggests that beyond the transition the phase is more tetragonal-like.

We also probed this phase transition using Raman spectroscopy. Figure $\ref{raman}$a shows a representative wavelength region of the Raman spectrum, between 190 and 340 cm$^{-1}$, for different temperatures. The intensity ratio between the phonon bands denoted A and B is changing as a function of temperature and exhibits a jump at a critical temperature of 380 K (Fig. \ref{raman}b). This change cannot be attributed to a phase transformation between the T-like and R-like phases as MS showed that their volume ratio remains constant with temperature. Rather, it reflects the phase transition occurring at $\sim$380 K as already found by XRD.

\begin{figure}
\includegraphics[width=0.85\columnwidth]{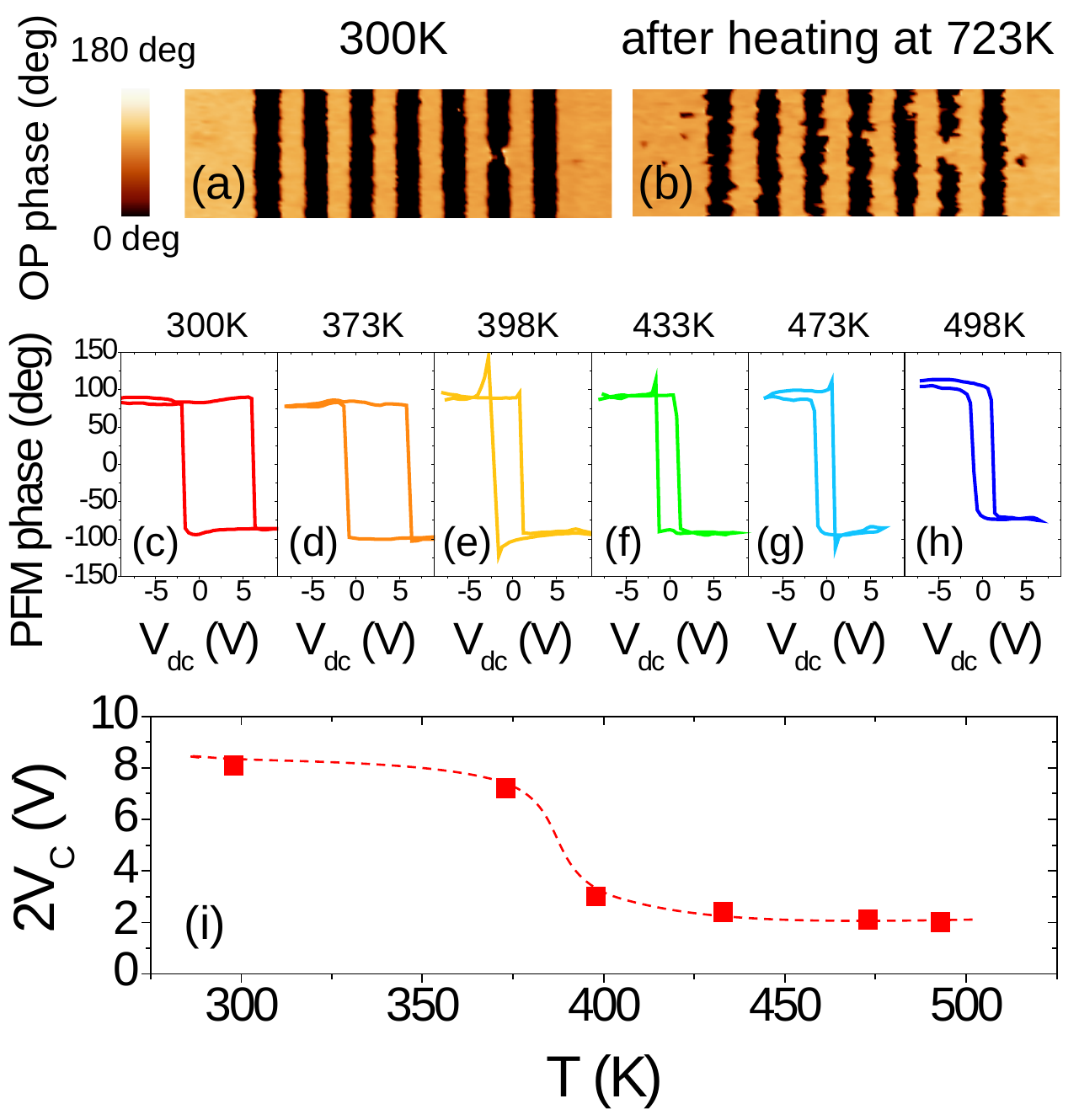}
\caption{(color online) (a-b) PFM out-of-plane phase images of artificial stripe domains written in a BFO/LSMO//LAO film at RT before (a) and after (b) heating up to 723 K. (c-g) Selected piezoresponse phase \emph{vs.} voltage cycles measured in a BFO/LSMO//LAO film  at different temperatures. (i) Temperature dependence of the coercive voltages extracted from the loops in (c-h). The dashed line is a guide to the eye.}
\label{pfm}
\end{figure}

To explore the FE nature of this phase transition we characterized the FE properties using PFM. We first wrote artificial stripe domains at RT and subsequently heated the sample \emph{ex situ} up to 723K during 2 hours. After cooling to 300 K, these domains could be imaged again (Fig. $\ref{pfm}$a and b), indicating that ferroelectricity is maintained up to at least 723K. We however note that the contrast deterioration at the domain edges may indicate that the transition to a non-FE state is near (see also \cite{infante2010}), consistent with the anomaly visible at $\sim$700 K in Fig. \ref{struc}a). Piezoelectric loops were also recorded while heating the samples in situ up to 498 K. At RT, loops can consistently be measured irrespective of the tip position on the sample, as expected for a standard FE, see Fig. $\ref{pfm}$c. However, at high temperatures, the response shows a strong variability with tip position, \emph{i.e.} either loops with low coercive field values (Fig. $\ref{pfm}$e-h) can be recorded or there is no response at all. This suggests that BFO//LAO exhibits a transition toward a phase with a soft FE response \cite{tan97} (visible on dependence of the coercive voltage with temperature, see Fig. $\ref{pfm}$h). Clearly this transition occurs in the same temperature range as the detected magnetic and structural phase transitions.

In summary, we have found that in BFO films deposited onto LAO substrate a majority T-like phase and a minority R-like one coexist. We argued that such phase coexistence allows to release the huge misfit strain imposed by the substrate. Remarkably, the T-like phase exhibits both structural, magnetic and FE phase transitions in a narrow temperature range around 360 K, as evidenced by XRD, M\"ossbauer spectroscopy and PFM studies, respectively, and confirmed by theory. Our results thus demonstrate the possibility of engineering a multiferroic having its critical temperatures (and thus diverging dielectric and magnetic susceptibilities, piezoelectricy, magnetoelectricity, etc) close to 300 K, which opens new paths for the use of BFO in applications.

\begin{acknowledgments}
We thank D. Colson, B. Warot-Fonrose, H. B\'ea, M.Couillard and G. Botton. This work was supported by the French C-Nano Ile-de-France Magellan, PRES Universud Nano-Ox, ANR Pnano project "M\'elo\"{\i}c", MICINN-Spain Grants No. MAT2010-18113, MAT2010-10093-E, CSD2007-00041, the ONR Grants No. N00014-04-1-0413, N00014-08-1- 0915, and N00014-07-1-0825, NSF Grants No. DMR 0701558 and DMR-0080054, and the Department of Energy, Office of Basic Energy Sciences, under contract ER-466120. Part of the microscopy work was carried out at the Canadian Centre for Electron Microscopy, a National Facility supported by NSERC and McMaster University.
\end{acknowledgments}


\begin{thebibliography}{23}
\expandafter\ifx\csname natexlab\endcsname\relax\def\natexlab#1{#1}\fi
\expandafter\ifx\csname bibnamefont\endcsname\relax
  \def\bibnamefont#1{#1}\fi
\expandafter\ifx\csname bibfnamefont\endcsname\relax
  \def\bibfnamefont#1{#1}\fi
\expandafter\ifx\csname citenamefont\endcsname\relax
  \def\citenamefont#1{#1}\fi
\expandafter\ifx\csname url\endcsname\relax
  \def\url#1{\texttt{#1}}\fi
\expandafter\ifx\csname urlprefix\endcsname\relax\def\urlprefix{URL }\fi
\providecommand{\bibinfo}[2]{#2}
\providecommand{\eprint}[2][]{\url{#2}}

\bibitem[{\citenamefont{Wang et~al.}(2009)\citenamefont{Wang, Liu, and
  Ren}}]{wang2009a}
\bibinfo{author}{\bibfnamefont{K.}~\bibnamefont{Wang}},
  \bibinfo{author}{\bibfnamefont{J.-M.} \bibnamefont{Liu}}, \bibnamefont{and}
  \bibinfo{author}{\bibfnamefont{Z.}~\bibnamefont{Ren}}, \bibinfo{journal}{Adv.
  Phys.} \textbf{\bibinfo{volume}{58}}, \bibinfo{pages}{321}
  (\bibinfo{year}{2009}).

\bibitem[{\citenamefont{B\'ea et~al.}(2008)\citenamefont{B\'ea, Gajek, Bibes,
  and Barth\'el\'emy}}]{bea2008a}
\bibinfo{author}{\bibfnamefont{H.}~\bibnamefont{B\'ea}},
  \bibinfo{author}{\bibfnamefont{M.}~\bibnamefont{Gajek}},
  \bibinfo{author}{\bibfnamefont{M.}~\bibnamefont{Bibes}}, \bibnamefont{and}
  \bibinfo{author}{\bibfnamefont{A.}~\bibnamefont{Barth\'el\'emy}},
  \bibinfo{journal}{J. Phys.: Condens. Matter} \textbf{\bibinfo{volume}{20}},
  \bibinfo{pages}{434221} (\bibinfo{year}{2008}).

\bibitem[{\citenamefont{Bibes et~al.}(2011)\citenamefont{Bibes, {J.E.
  Villegas}, and Barth\'el\'emy}}]{bibes2011}
\bibinfo{author}{\bibfnamefont{M.}~\bibnamefont{Bibes}},
  \bibinfo{author}{\bibnamefont{{J.E. Villegas}}}, \bibnamefont{and}
  \bibinfo{author}{\bibfnamefont{A.}~\bibnamefont{Barth\'el\'emy}},
  \bibinfo{journal}{Adv. Phys.} \textbf{\bibinfo{volume}{60}},
  \bibinfo{pages}{5} (\bibinfo{year}{2011}).

\bibitem[{\citenamefont{Catalan and {J.F. Scott}}(2009)}]{catalan2009}
\bibinfo{author}{\bibfnamefont{G.}~\bibnamefont{Catalan}} \bibnamefont{and}
  \bibinfo{author}{\bibnamefont{{J.F. Scott}}}, \bibinfo{journal}{Adv. Mater.}
  \textbf{\bibinfo{volume}{21}}, \bibinfo{pages}{2463} (\bibinfo{year}{2009}).

\bibitem[{\citenamefont{{I.C. Infante} et~al.}(2010)}]{infante2010}
\bibinfo{author}{\bibnamefont{{I.C. Infante}}} \bibnamefont{et~al.},
  \bibinfo{journal}{Phys. Rev. Lett.} \textbf{\bibinfo{volume}{105}},
  \bibinfo{pages}{057601} (\bibinfo{year}{2010}).

\bibitem[{\citenamefont{{A.J. Hatt.} et~al.}(2010)\citenamefont{{A.J. Hatt.},
  {N.A. Spaldin}, and Ederer}}]{hatt2010}
\bibinfo{author}{\bibnamefont{{A.J. Hatt.}}},
  \bibinfo{author}{\bibnamefont{{N.A. Spaldin}}}, \bibnamefont{and}
  \bibinfo{author}{\bibfnamefont{C.}~\bibnamefont{Ederer}},
  \bibinfo{journal}{Phys. Rev. B} \textbf{\bibinfo{volume}{81}},
  \bibinfo{pages}{054109} (\bibinfo{year}{2010}).

\bibitem[{\citenamefont{Di\'eguez et~al.}(2011)\citenamefont{Di\'eguez, {O.E.
  Gonzalez-Vazquez}, {J.C. Wojdel}, and {J. I\~niguez}}}]{dieguez2011}
\bibinfo{author}{\bibfnamefont{O.}~\bibnamefont{Di\'eguez}},
  \bibinfo{author}{\bibnamefont{{O.E. Gonzalez-Vazquez}}},
  \bibinfo{author}{\bibnamefont{{J.C. Wojdel}}}, \bibnamefont{and}
  \bibinfo{author}{\bibnamefont{{J. I\~niguez}}}, \bibinfo{journal}{Phys. Rev.
  B} \textbf{\bibinfo{volume}{83}}, \bibinfo{pages}{0940105}
  (\bibinfo{year}{2011}).

\bibitem[{\citenamefont{Ricinschi et~al.}(2006)\citenamefont{Ricinschi, {K.-Y.
  Yun}, and {M.A. Okuyama}}}]{ricinschi2006}
\bibinfo{author}{\bibfnamefont{D.}~\bibnamefont{Ricinschi}},
  \bibinfo{author}{\bibnamefont{{K.-Y. Yun}}}, \bibnamefont{and}
  \bibinfo{author}{\bibnamefont{{M.A. Okuyama}}}, \bibinfo{journal}{J. Phys.:
  Condens. Matter.} \textbf{\bibinfo{volume}{18}}, \bibinfo{pages}{L97}
  (\bibinfo{year}{2006}).

\bibitem[{\citenamefont{B\'ea et~al.}(2009)}]{bea2009}
\bibinfo{author}{\bibfnamefont{H.}~\bibnamefont{B\'ea}} \bibnamefont{et~al.},
  \bibinfo{journal}{Phys. Rev. Lett.} \textbf{\bibinfo{volume}{102}},
  \bibinfo{pages}{217603} (\bibinfo{year}{2009}).

\bibitem[{\citenamefont{{H.M. Christen} et~al.}(2011)\citenamefont{{H.M.
  Christen}, {J.H. Nam}, {H.S. Kim}, {A.J. Hatt}, and {N.A.
  Spaldin}}}]{christen2011}
\bibinfo{author}{\bibnamefont{{H.M. Christen}}},
  \bibinfo{author}{\bibnamefont{{J.H. Nam}}},
  \bibinfo{author}{\bibnamefont{{H.S. Kim}}},
  \bibinfo{author}{\bibnamefont{{A.J. Hatt}}}, \bibnamefont{and}
  \bibinfo{author}{\bibnamefont{{N.A. Spaldin}}}, \bibinfo{journal}{Phys. Rev.
  B} \textbf{\bibinfo{volume}{83}}, \bibinfo{pages}{144107}
  (\bibinfo{year}{2011}).

\bibitem[{\citenamefont{{R.J. Zeches} et~al.}(2009)}]{zeches2009}
\bibinfo{author}{\bibnamefont{{R.J. Zeches}}} \bibnamefont{et~al.},
  \bibinfo{journal}{Science} \textbf{\bibinfo{volume}{326}},
  \bibinfo{pages}{977} (\bibinfo{year}{2009}).

\bibitem[{\citenamefont{{K.J. Choi} et~al.}(2004)}]{choi2004}
\bibinfo{author}{\bibnamefont{{K.J. Choi}}} \bibnamefont{et~al.},
  \bibinfo{journal}{Science} \textbf{\bibinfo{volume}{306}},
  \bibinfo{pages}{1005} (\bibinfo{year}{2004}).

\bibitem[{\citenamefont{{J.C. Wojdel} and {J. I\~niguez}}(2010)}]{wojdel2010}
\bibinfo{author}{\bibnamefont{{J.C. Wojdel}}} \bibnamefont{and}
  \bibinfo{author}{\bibnamefont{{J. I\~niguez}}}, \bibinfo{journal}{Phys. Rev.
  Lett.} \textbf{\bibinfo{volume}{105}}, \bibinfo{pages}{037208}
  (\bibinfo{year}{2010}).

\bibitem[{\citenamefont{Prosandeev et~al.}(2011)\citenamefont{Prosandeev, {I.A.
  Kornev}, and Bellaiche}}]{prosandeev2011}
\bibinfo{author}{\bibfnamefont{S.}~\bibnamefont{Prosandeev}},
  \bibinfo{author}{\bibnamefont{{I.A. Kornev}}}, \bibnamefont{and}
  \bibinfo{author}{\bibfnamefont{L.}~\bibnamefont{Bellaiche}},
  \bibinfo{journal}{Phys. Rev. B} \textbf{\bibinfo{volume}{83}},
  \bibinfo{pages}{020102(R)} (\bibinfo{year}{2011}).

\bibitem[{\citenamefont{B\'ea et~al.}(2005)}]{bea2005}
\bibinfo{author}{\bibfnamefont{H.}~\bibnamefont{B\'ea}} \bibnamefont{et~al.},
  \bibinfo{journal}{Appl. Phys. Lett.} \textbf{\bibinfo{volume}{87}},
  \bibinfo{pages}{072508} (\bibinfo{year}{2005}).

\bibitem[{\citenamefont{Juraszek et~al.}(2009)\citenamefont{Juraszek, Zivotsky,
  Chiron, Vaudolon, and Teillet}}]{juraszek2009}
\bibinfo{author}{\bibfnamefont{J.}~\bibnamefont{Juraszek}},
  \bibinfo{author}{\bibfnamefont{O.}~\bibnamefont{Zivotsky}},
  \bibinfo{author}{\bibfnamefont{H.}~\bibnamefont{Chiron}},
  \bibinfo{author}{\bibfnamefont{C.}~\bibnamefont{Vaudolon}}, \bibnamefont{and}
  \bibinfo{author}{\bibfnamefont{J.}~\bibnamefont{Teillet}},
  \bibinfo{journal}{Rev. Sci. Instrum.} \textbf{\bibinfo{volume}{80}},
  \bibinfo{pages}{043905} (\bibinfo{year}{2009}).

\bibitem[{\citenamefont{Dup\'e et~al.}(2010)}]{dupe2010}
\bibinfo{author}{\bibfnamefont{B.}~\bibnamefont{Dup\'e}} \bibnamefont{et~al.},
  \bibinfo{journal}{Phys. Rev. B} \textbf{\bibinfo{volume}{81}},
  \bibinfo{pages}{144128} (\bibinfo{year}{2010}).

\bibitem[{\citenamefont{Chen et~al.}(2011)}]{chen2011}
\bibinfo{author}{\bibfnamefont{Z.}~\bibnamefont{Chen}} \bibnamefont{et~al.}
  (\bibinfo{year}{2011}), \bibinfo{note}{arXiv.org:1104.4712v1}.

\bibitem[{\citenamefont{Di\'eguez and {J. I\~niguez}}(2011)}]{dieguez2011b}
\bibinfo{author}{\bibfnamefont{O.}~\bibnamefont{Di\'eguez}} \bibnamefont{and}
  \bibinfo{author}{\bibnamefont{{J. I\~niguez}}} (\bibinfo{year}{2011}),
  \bibinfo{note}{arXiv.org:1105.2043}.

\bibitem[{\citenamefont{Toupet et~al.}(2010)\citenamefont{Toupet, {F. Le
  Marrec}, Lichtensteiger, Dkhil, and Karkut}}]{toupet2010}
\bibinfo{author}{\bibfnamefont{H.}~\bibnamefont{Toupet}},
  \bibinfo{author}{\bibnamefont{{F. Le Marrec}}},
  \bibinfo{author}{\bibfnamefont{C.}~\bibnamefont{Lichtensteiger}},
  \bibinfo{author}{\bibfnamefont{B.}~\bibnamefont{Dkhil}}, \bibnamefont{and}
  \bibinfo{author}{\bibfnamefont{M.}~\bibnamefont{Karkut}},
  \bibinfo{journal}{Phys. Rev. B} \textbf{\bibinfo{volume}{81}},
  \bibinfo{pages}{140101(R)} (\bibinfo{year}{2010}).

\bibitem[{\citenamefont{Haumont et~al.}(2008)}]{haumont2008}
\bibinfo{author}{\bibfnamefont{R.}~\bibnamefont{Haumont}} \bibnamefont{et~al.},
  \bibinfo{journal}{Phys. Rev. B} \textbf{\bibinfo{volume}{78}},
  \bibinfo{pages}{134108} (\bibinfo{year}{2008}).

\bibitem[{\citenamefont{Noheda et~al.}(2002)}]{noheda2002}
\bibinfo{author}{\bibfnamefont{N.}~\bibnamefont{Noheda}} \bibnamefont{et~al.},
  \bibinfo{journal}{Phys. Rev. B} \textbf{\bibinfo{volume}{65}},
  \bibinfo{pages}{224101} (\bibinfo{year}{2002}).

\bibitem[{\citenamefont{Tan et~al.}(1997)\citenamefont{Tan, {J.F. Li}, and
  Viehland}}]{tan97}
\bibinfo{author}{\bibfnamefont{Q.}~\bibnamefont{Tan}},
  \bibinfo{author}{\bibnamefont{{J.F. Li}}}, \bibnamefont{and}
  \bibinfo{author}{\bibfnamefont{D.}~\bibnamefont{Viehland}},
  \bibinfo{journal}{Philos. Mag. B} \textbf{\bibinfo{volume}{76}},
  \bibinfo{pages}{59} (\bibinfo{year}{1997}).

\end{thebibliography}
\end{document}